\documentclass[cup6b,epsf]{cupbook}

\usepackage{epsf}

\newcommand{\modot}{M$_\odot$\hspace*{0.05cm}}
\def\sles{\lower2pt\hbox{$\buildrel {\scriptstyle <}
   \over {\scriptstyle\sim}$}}
\def\sgreat{\lower3pt\hbox{$\buildrel {\scriptstyle >}
   \over {\scriptstyle\sim}$}}

\begin{document}

\author{Adam Burrows\\
Department of Astronomy and \\
Steward Observatory, The University of Arizona, Tucson, AZ 85721 \and
Christian D. Ott\\
Institut f\"{u}r Theoretische Astrophysik, Universit\"{a}t Heidelberg \and
Casey Meakin\\
Department of Astronomy and \\
Steward Observatory, The University of Arizona, Tucson, AZ 85721}

\chapter{Topics in Core-Collapse Supernova Theory}

\vspace{0.2in}
{\it There are many interesting topics at the intersection of physics and astrophysics
we call Supernova Theory.  A small subset of them include the origin of pulsar kicks, 
gravitational radiation signatures of core bounce, and the possible roles of neutrinos and rotation in the mechanism
of explosion.  In this brief communication we summarize various recent ideas and calculations
that bear on these themes.    
}

\section{What is the Mechanism of Pulsar Kicks?}
\label{kick}

Radio pulsars are observed to have large proper motions that average
$\sim$400-500 km s$^{-1}$ (Lyne \& Lorimer 1994) and whose velocity distribution might be bimodal 
(Fryer, Burrows, and Benz 1998; Arzoumanian, Chernoff, \& Cordes 2002).
If bimodal, the slow peak would have a mean speed near $\sim$100 km s$^{-1}$
and the fast peak would have a mean speed near 500--600 km s$^{-1}$.
A bimodal distribution implies different populations and different mechanisms, but what these
populations could be remains highly speculative.

Many arguments suggest that pulsars are given ``kicks"
at birth (Lai 2000; Lai, Chernoff, and Cordes 2001), and 
are not accelerated over periods of years or centuries. 
The best explanation is that these kicks are imparted
during the supernova explosion itself.  We think that this view 
is compelling.  The two suggested modes of acceleration and impulse 
are via net neutrino anisotropy during the neutrino emission
phase (which lasts seconds) and anisotropic mass motions and aspherical explosion which impart
momentum to the residual core.  The former requires but a $\sim$1\% {\it net}
anisotropy in the neutrino angular distribution to provide a $\sim$300 km s$^{-1}$
kick.  However, anisotropies in the neutrino radiation field are more
easily smoothed than matter anisotropies due to convection, rotation,
aspherical collapse, etc. and relativistic particles such as neutrinos
are not as efficient as non-relativistic matter at converting a
given amount of energy into recoil (momentum).  To achieve the requisite
neutrino anisotropies people have generally invoked large ($\sim$10$^{15-16}$ gauss)
magnetic fields, which may not obtain generically (see Lai 2000 for a summary).  Furthermore, all
multi-D calculations to date imply that convective motions between
the inner core and the shock result in significant jostling of
the protoneutron star.  Velocities of $\sim$100-200 km s$^{-1}$
arise quite naturally by dint of the basic hydrodynamics 
of the convective mantle of the iron core after bounce
(``Brownian Motion"; Burrows, Hayes, and Fryxell 1995 (BHF);
Burrows and Hayes 1996a; Janka and M\"uller 1994; Scheck et al. 2003).  This process {\it must} 
be a stochastic contributor to pulsar proper motions.  In addition, 
due to the associated torques, modest spins can be imparted (Burrows, Hayes, and Fryxell 1995).

However, the average recoil speeds obtained theoretically by Burrows, Hayes, and Fryxell (1995),
Burrows and Hayes (1996a), and Scheck et al. (2003) due to the brownian motion of the core
are only $\sim$200 km s$^{-1}$; this is not sufficient to explain either
the average pulsar speed or the high-speed peak of a bimodal distribution.  
To do that might require an initial mild anisotropy ($\sim$percents) in the 
density or velocity profiles of the collapsing Chandrasekhar core (Lai 2000; Lai and Goldreich 2000).  Such
small anisotropies have been shown to result in significant impulses and implied kicks of 550 to 800 km s$^{-1}$
(Burrows and Hayes 1996b).  It may be that whatever determines whether the
initial core is anisotropic results in the high-speed peak of the bimodal
distribution, while the low-speed peak is due to the natural jostling
by convective plumes and the resultant brownian motion of the core.
The latter is stochastic and not deterministic, but has been a robust
prediction of the collapse theory for many years.  

An instability in the 
pre-collapse structure that might result in aspherical collapse (particularly
relevant in the supersonic region of the collapse since it can not smooth
itself out by pressure forces) might be progenitor mass dependent;  high-mass
progenitors might result in high-velocity pulsars, while low-mass progenitors
might result in low-velocity pulsars (on average) (or vice versa!).  Whatever the origin
of pulsar kicks and their apparent bimodality may be, new calculations are desparately
needed.  No hydrodynamic calculation to date has actually freed the very inner core
to respond to the pressure impulses in a consistent fashion.  All calculations have
anchored the core and recoils have been inferred due to the integrated anisotropic pressure distributions
seen.  Freeing the core to respond to pressure and gravity effects and allowing 
the associated feedback processes will be crucial for obtaining
self-consistent and credible results.

\section{Gravitational Waves from Core Collapse}
\label{grav}

Gravitational radiation signatures can in 
principle provide a dramatic potential constraint
on core-collapse supernovae. 
Massive stars (ZAMS mass $\sgreat$ 8 \modot) develop degenerate
cores in the final stages of nuclear burning and achieve the Chandrasekhar
mass.  Gravitational collapse ensues, leading to dynamical compression to nuclear
densities, subsequent core bounce, and hydrodynamical shock wave generation.
These phenomena involve large masses at high velocities ($\sim c/4$) and great
accelerations.  Such dynamics, if only slightly aspherical, will lead
to copious gravitational wave emission and, arguably, to one of the most
distinctive features of core-collapse supernovae. The gravitational waveforms and
associated spectra bear the direct stamp of the hydrodynamics and rotation
of the core and speak volumes about internal supernova evolution.
Furthermore, they provide data that complement (temporally
and spectrally) those from the neutrino pulse (which also originates from the core), enhancing
the diagnostic potential of each.

Most stars rotate and rotation can result in large asphericity at and around bounce.
This provides hope that the emission of gravitational radiation from stellar
core collapse can be significant.  Furthermore, Rayleigh-Taylor-like convection in the protoneutron
star, the aspherical emission of neutrinos, and post-bounce triaxial rotational instabilities
are also potential sources of gravitational radiation.
Together these phenomena, with their characteristic spectral and temporal signatures,
make core-collapse supernovae promising and interesting generators of gravitational radiation.

Ott et al. (2003) use the 2D hydro code VULCAN/2D (Livne 
1993) and follow Zwerger \& M\"uller (1997) in forcing 
the one-dimensional initial models to rotate with constant
angular velocity on cylinders according to the rotation law
\begin{equation}
\label{eq:rotlaw}
\Omega(r) = \Omega_0 \, \bigg[ 1 + \bigg(\frac{r}{{\rm A}}\bigg)^2 \bigg]^{-1}\, ,
\label{omegaeq}
\end{equation}
where $\Omega(r)$ is the angular velocity, $r$ is the distance from
the rotation axis, and $\Omega_0$ and A are free parameters that
determine the rotational speed/energy of the model and the scale of the 
distribution of angular momentum.  The rotation parameter $\beta$ is 
defined by
\begin{equation}
\beta = \frac{E_{rot}}{|E_{grav}|}\, ,
\end{equation}
where $E_{rot}$ is the total rotational kinetic energy and $E_{grav}$
is the total gravitational energy.
We (Ott et al. 2003) name our runs according to the following convention:
[initial model name]{\rm A}[in km]$\beta_i$[in \%]. For example, s11A1000$\beta$0.3 is a Woosley
and Weaver (1995) 11 \modot model with A=1000 km and an initial $\beta_i$ of 0.3\%.

Representative results are those found 
for model s15A1000$\beta$0.2 (Ott et al. 2003).
The spectrum of s15A1000$\beta$0.2 
is dominated by frequencies between 300 Hz and 600 Hz and peaks at 460~Hz. Most
of the smaller peaks are connected to the first spike in the waveform during
which 94\% of the total gravitational wave energy of this model is radiated.
There is, however, a contribution by the
radial and non-radial ring-down pulsations that have
characteristic periods of 2 - 2.5 ms in this 
model, translating into frequencies of 400-500~Hz.
The peak is at 700 Hz and there are higher 
harmonics around 1400 Hz. With increasing $\beta_i$
the spectrum shifts to lower frequencies and lower absolute values,
peaking at 152 Hz ($\beta_i$=0.40\%),
91 Hz ($\beta_i$=0.60\%), and 38 Hz ($\beta_i$=0.80\%). Furthermore, a prominent peak
at low frequencies can be directly associated with the oscillation
frequency of the post bounce cycles.

The models of Ott et al. (2003) yield absolute values of 
the dimensionless maximum gravitational wave strain
in the interval 2.0 $\times$ 10$^{-23}$ $\le$ $h^{TT}_{max}$ $\le$
1.25 $\times$ 10$^{-20}$ at a distance of 10 kpc.
The total energy radiated ($E_{GW}$) lies
in the range 1.4 $\times$ 10$^{-11}$ \modot c$^2$ $\le$ $E_{GW}$ $\le$
2.21 $\times$ 10$^{-8}$ \modot c$^2$ and the energy spectra peak (with the exception
of a very few models) in the frequency interval 20 Hz \sles$\,$ f$_{peak}$ \sles$\,$~600~Hz.

\begin{figure}
\begin{center}
\leavevmode\epsfxsize=10cm \epsfbox{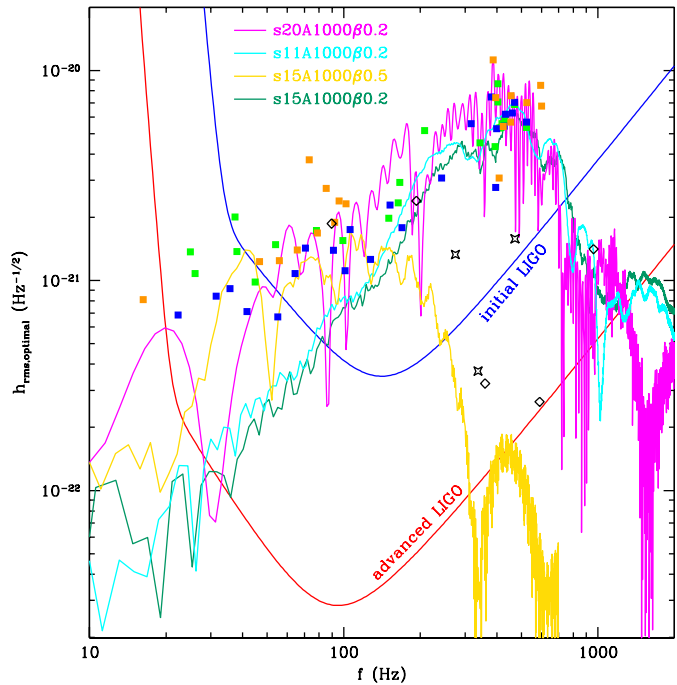}
\end{center}
\caption{
LIGO sensitivity plot. Plotted are the optimal root-mean-square noise
strain amplitudes $h_{rms} = \sqrt{f S(f)}$ of the initial and advanced LIGO
interferometer designs. Optimal means that the gravitational waves are
incident at an optimal angle and optimal polarization for detection and that there
are coincident measurements of gravitational waves by
multiple detectors. For gravitational waves from
burst sources incident at random times from a random direction and a signal-to-noise
ratio (SNR) of 5, the rms noise level $h_{rms}$ is approximately a factor of 11 above
the one plotted here (Abramovici et al. 1992)
We have plotted solid squares at the maxima of the characteristic gravitational wave
strain spectrum ($h_{char} (f)$) of
our s11, s15, and s20 models from Woosley and Weaver (1995) that
were artificially put into
rotation. Our nonrotating models are marked with stars; diamonds stand for models
from Heger et al. (2000, 2003). The distance to Earth was set to 10 kpc for
all models. Most of our models lie above the optimal design sensitivity limit of LIGO I.
Hence, the prospects for detection are good.
Those models that are not detectable by the 1st-generation LIGO
are those that rotate most slowly
and those which are the fastest rotators. See Ott et al. (2003) for details.
}
\label{fig:grav}
\end{figure}

Ott et al. (2003) find that at a distance of 10 kpc, i.e. for galactic
distances, the 1st-generation LIGO, once it has reached its design sensitivity level,
will be able to detect more than 80\% of our core collapse
models under optimal conditions and orientations.
Assuming random polarizations and angles of incidence, this reduces to 10\%.
Advanced LIGO, however, should be able to detect virtually all models at galactic distances.
Figure \ref{fig:grav} presents peak $h_{char}$ (the points), the maxima of the characteristic gravitational wave
strain spectrum, but it also includes the actual $h_{char}$ 
spectra of selected models (see Ott et al. 2003 for details).
These $h_{char}$ serve to put the issues of detectability in the LIGO detector into sharp relief.

\section{Rotational Effects}
\label{rot}

The evolution of the rotation parameter $\beta$
and of the angular velocity is of particular interest, since they are connected
to two still unanswered questions in core-collapse supernovae physics: What are
the periods of newborn neutron stars? 
What is the role of rotation in the mechanism of core-collapse supernovae?
As a prelude, Ott et al. (2003) addressed two related points:

1) There exists a maximum
value of $\beta$ at bounce for a given progenitor model and value of A. Interestingly,
the maximum $\beta$ is not reached by the model with the maximum $\beta_i$, but
by a model with some intermediate value of $\beta_i$.
$\beta$ at bounce is determined by the subtle interplay between initial angular
momentum distribution, the equation of state,
centrifugal forces and gravity. The ``optimal'' configuration
leads to the overall maximum $\beta$ at bounce for a given $\beta_i$.
Generally, $\beta$ increases during collapse by a factor of $\sim$10-40.

2) As with $\beta$, overall the angular velocity increases with
increasing $\beta_i$ until a maximum is reached. It subsequently decreases
with the further increase of $\beta_i$. The initially more
rigidly rotating models actually yield larger
post-bounce angular velocity gradients inside 30 km.
The equatorial velocity profile peaks off center for moderate
$\beta_i$ at radii between 6 and 10 km.  An initially more differentially rotating
model (at a given $\beta_i$) leads to the highest central values of the
angular velocity, while its angular velocity profile quickly drops to low
values and near rigid rotation for $\beta_i \ge$ 0.3\%.
Model s15A500$\beta$0.2 in Ott et al. (2003) (see \S\ref{grav}) results
in the shortest rotation period near the center ($\sim$1.5 ms).
Model s15A50000$\beta$0.5 yields the 
shortest period of the A=50000 km model series ($\sim$1.85 ms).

In sum, the amplification of the angular velocity (frequency) due to collapse is generally
large, from a factor of $\sim$25 to $\sim$1000.  An initial period of 2 seconds
in the iron core can translate into a period at bounce 
of $\sim$5 {\it milliseconds}, depending upon
the initial rotational profile.  
The angular velocity shear exterior to the peak at 6-10 km
exhibited by these models has also been identified in the one-dimensional
study of Akiyama et al. (2003). These authors
consider such shear a possible driver for the magneto-rotational
instability (MRI), which could be a generator of strong magnetic fields.

\subsection{Rotation and Explosion}

The large amplification of the angular velocity during bounce implies
that rotation may be a factor in core collapse phenomenology and in the 
explosion mechanism.  Though the latter remains to be demonstrated, there
are a few aspects of rotating collapse that bear mentioning and that distinguish it from
spherical collapse:  1) Rotation lowers the effective
gravity in the core, increasing the radius of the stalled shock and the 
size of the gain region.  Since ejection is inhibited by the deep
potential well, rotation might in this manner facilitate explosion.
2) Rotation generates vortices that might dredge up heat from 
below the neutrinospheres and thereby enhance the driving neutrino luminosities.
3) Rotation lowers (slightly) the optical depth of a given mass shell,
thereby increasing the $\nu_e$ neutrino luminosity.  (However, as Fryer and 
Heger (2000) have shown the $\bar{\nu}_e$ luminosity is at the same time decreased due
to the lower temperatures achieved.) 4) Importantly, rotation results
in a pronounced anisotropy in the mass accretion flux after bounce.  In
fact, rotation can create large pole-to-equator differences in the
density profiles of the infalling matter, due to the centrifugal barrier
along the poles.  The actual magnitude and evolution
of this barrier is a function of the degree of rotation and its profile,
but can be quite pronounced.  Very approximately, in the equatorial region the 
distance from the axis ($\rho$, in cylindrical coordinates) of the barrier
is given by $j^2/(GM)$, where $M$ is the interior mass and $j$ is the specific
angular momentum at that mass.  If the slope of $j$ with $r$ is positive, then
as matter from further and further out accretes onto 
the protoneutron star the centrifugal barrier is 
expected to grow in extent.  Even if the $j$ profile is flat, $j^2/(GM)$
might be an interesting ($\sim$10-300 kilometers ?) number (Heger et al. 2000,2003).

\begin{figure}
\begin{center}
\leavevmode\epsfxsize=10cm \epsfbox{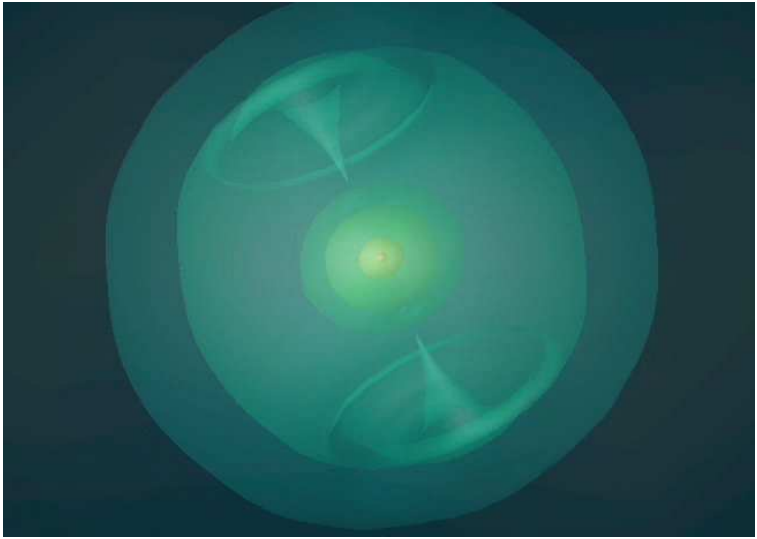}
\end{center}
\caption{A snapshot of a rotating collapse and bounce simulation in 2D,
rendered in 3D with nested layers of isodensity contours from 10$^8$ gm cm$^{-3}$
to 10$^{13}$ gm cm$^{-3}$.  The funnel along the poles due to the 
centrifugal barrier created after bounce by the rotation of the collapsing
Chandrasekhar core is clearly seen.
}
\label{funnels}
\end{figure}

Figure \ref{funnels} depicts a snapshot
after bounce of various nested isodensity contours for a rapidly
rotating initial progenitor ($\Omega_0$ $\sim$ $2\pi/2$ rad s$^{-1}$; eq. \ref{omegaeq}).  The fact that outer
(lower-density) contours pierce into the inner regions along the poles
implies that a partially evacuated region is carved out along these poles.
Since the total accretion rate is set by the initial mass density profile
and since accretion powers the early post-bounce neutrino luminosity, this luminosity
is not much changed.  However, due to significant rotation the mass accretion flux at the poles is
small after bounce.  This suggests that the neutrino-driven mechanism is naturally
facilitated along the poles (Burrows \& Goshy 1993).  A 
bipolar explosion would result (Fryer \& Heger 2000,
though see Buras et al. 2003).
Thus, such bipolarity (and the consequent optical polarization    
of the debris) is not an exclusive signature of MHD driven explosions
and may be a natural consequence of the neutrino-driven mechanism with rotation. 
However, for rotation to be pivotal in the mechanism, it might need to be rapid.  
This would beg the question of how the excess angular momentum
is ejected to leave only the modestly rotating pulsars observed.  Clearly, much works 
remains to be done.

\section{Reprise on Supernova Energetics, Made simple}

The discussion concerning the rudiments of supernova energetics
included in Burrows and Thompson (2002) summarizes our thoughts on the
origin of the supernova energy scale and contains a useful perspective on the
true efficiency of the neutrino-driven mechanism.  It is reprised here
without shame due to the continuing interest outside the supernova community
in simple arguments that are not as opaque as numerical simulations can be
(Thompson, Burrows, and Pinto 2003; Liebend\"orfer et al. 2001; Rampp \& Janka 2000).

\subsection{Supernova Energetics Made Simple (?)}

``It is important to note that one is not obliged to unbind the inner core ($\sim$10 kilometers) as well; the 
explosion is a phenomenon of the outer mantle at ten times the radius (50-200 kilometers).  
One consequence of this goes to the heart of a general confusion concerning 
supernova physics.  Though the binding energy of a cold neutron star is $\sim$$3 \times 10^{53}$ ergs
and the supernova explosion energy is near $10^{51}$ ergs, a comparison of these two numbers and
the large ratio that results are not very relevant.  More germane are the binding energy
of the mantle (interior to the shock or, perhaps, exterior to the neutrinospheres)
and the neutrino energy radiated during 
the delayed phase.  These are both at most a few$\times$10$^{52}$ ergs, not
$\sim$$3 \times 10^{53}$ ergs, and the relevant ratio that illuminates
the neutrino-driven supernova phenomenon is $\sim$10$^{51}$ 
ergs divided by a few$\times$10$^{52}$ ergs. This is $\sim$5-10\%, not 
the oft-quoted 1\%, a number which tends to overemphasize the 
sensitivity of the neutrino mechanism to neutrino and numerical details.       

Furthermore, there is general confusion concerning what determines the supernova explosion energy.
While a detailed understanding of the supernova mechanism is required to
answer this question, one can still proffer a few observations.  First is the simple
discussion above.  Five to ten percent of the neutrino energy coursing through
the semi-transparent region is required, not one percent.  Importantly, the optical
depth to neutrino absorption in the gain region is of order $\sim$0.1.  The product
of the sum of the $\nu_{e}$ and $\bar{\nu}_e$ neutrino energy emissions in the first
100's of milliseconds and this optical depth gives a number near 10$^{51}$ ergs.
Furthermore, the binding energy of the progenitor mantle exterior to the iron core
is of order a few$\times 10^{50}$ to a few$\times 10^{51}$ ergs and it is very approximately
this binding energy, not that of a cold neutron star, that is relevant in setting the 
scale of the core-collapse supernova explosion energy.  Given the power-law nature
of the progenitor envelope structure, it is clear that this binding energy is related
to the binding energy of the pre-collapse iron core (note that they both have a boundary
given by the same $GM/R$), which at collapse is that of
the Chandrasekhar core.  The binding energy of the Chandrasekhar core is easily
shown to be zero, modulo the rest mass of the electron times the number of baryons
in a $\sim$1.4 M$_{\odot}$ Chandrasekhar mass.  (The Chandrasekhar mass/instability is tied to the 
onset of relativity for the electrons, itself contingent upon the electron rest mass).  
The result is $\sim$10$^{51}$ ergs.

The core-collapse explosion energy is near the explosion energy 
for a Type Ia supernovae because in a thermonuclear explosion
the total energy yield is approximately the 0.5 MeV/baryon derived from carbon/oxygen burning to
iron times the number of baryons burned in the explosion.  
The latter is $\ge$half the number of baryons in a Chandrasekhar
mass.  The result is $\sim$10$^{51}$ ergs.  This is the same number as for core-collapse supernovae because
1) in both cases we are dealing with the Chandrasekhar mass (corrected for electron captures,
entropy, general relativity, and Coulomb effects) and 2) the electron mass and the per-baryon
thermonuclear yield are each about 0.5 MeV.

While more detailed calculations are clearly necessary to do this correctly, the essential
elements of supernova energetics are not terribly esoteric (if 
neutrino-driven), at least to within a factor of 5,
and should not be viewed as such."

\bigskip\noindent
{\it Acknowledgements} Support for this work is provided in part by
the Scientific Discovery through Advanced Computing (SciDAC) program
of the DOE, grant number DE-FC02-01ER41184,
a NASA GSRP program fellowship,
and by NASA through Hubble Fellowship
grant \#HST-HF-01157.01-A awarded by the Space Telescope Science
Institute, which is operated by the Association of Universities for Research in Astronomy,
Inc., for NASA, under contract NAS 5-26555.


\begin{thereferences}{99}

\makeatletter
\renewcommand{\@biblabel}[1]{\hfill}

\bibitem[{Abramovici et al.}(1992)]{abra:92}
Abramovici, A. et al., 1992. {\it Science}, {\bf 256}, 325.
\bibitem[{Akiyama {et~al.}(2003){Akiyama} et al.}]{Akiyama:03}
Akiyama, S., Wheeler, J.C., Meier, D., \& Lichtenstadt, I.,
2003. {\it Astrophys.~J.}, {\bf 584}, 954.
\bibitem[]{}
Arzoumanian, Z., Chernoff, D.F., \&  Cordes, J., 2002.
{\it Astrophys. J.}, {\bf 568}, 289--301.
\bibitem[Buras et al.~(2003)]{buras}
Buras, R., Rampp, M., Janka, H.-Th., \& Kifonidis, K. 2003, {\it Phys. Rev. Lett.}, {\bf 90}, 241101.
\bibitem[]{bhf_1995}
Burrows, A., Hayes, J., \& Fryxell, B.A., 1995. {\it Astrophys.~J.}, {\bf 450}, 830.
\bibitem[]{} Burrows, A \& and Hayes, J., 1996a.
``An Origin for Pulsar Kicks in Supernova Hydrodynamics,'' p. 25, in
the Proceedings of the conference, {\it High-Velocity Neutron Stars and Gamma-Ray Bursts},
eds. R. E. Rothschild \& R. E. Lingenfelter,  A.I.P. Press, no. 366.
\bibitem[]{}
Burrows, A. \& Hayes, J., 1996b.
``Pulsar Recoil and Gravitational Radiation due to Asymmetrical Stellar Collapse and Explosion,''
{\it Phys. Rev. Letters}, {\bf 76}, 352.
\bibitem[]{}
Burrows, A. \& Goshy, J., 1993. {\it Astrophys.~/J.}, {\bf 416}, L75.
\bibitem[]{}
Burrows, A. \& Thompson, T.A., 2002.
``The Mechanism of Core-Collapse Supernova Explosions: A Status Report," 
in the proceedings of the ESO/MPA/MPE Workshop 
{\it From Twilight to Highlight: The Physics of
Supernovae}, p. 53, eds. Bruno Leibundgut and Wolfgang Hillebrandt (Springer-Verlag).
\bibitem[]{}
Fryer, C.L., Burrows, A., \& Benz, W., 1998. {\it Astrophys.~J.}, {\bf 496}, 333.
\bibitem[Fryer \& Heger (2000)]{fryerheger} 
Fryer, C.L. \& Heger, A., 2000. {\it Astrophys.~J.}, {\bf 541}, 1033.
\bibitem[{{Heger}, {Langer}, and {Woosley}(2000){Heger}, {Langer}, and {Woosley}}]{heger:00}
{Heger}, A., {Langer}, N., and {Woosley}, S.E., 2000. {\it Astrophys.~J.}, {\bf 528}, 368.
\bibitem[{Heger {et~al.}(2003)Heger, Woosley, Langer, and Spruit}]{spruit:03}
Heger, A., Woosley, S.E., Langer, N., \& Spruit, H.C., 2003. Stellar Rotation, 
Proceedings IAU Symposium No. 215.
\bibitem[Janka and M\"uller]{jm94}
Janka, H.-Th. \& M\"uller, E., 1994.
{\it Astron. and Astrophys.}, {\bf 290}, 496--502.
\bibitem[]{}
Lai, d. \& Goldreich, P., 2000.  {\it Astrophys.~J.}, {\bf 535}, 402.
\bibitem[]{}
Lai, D., 2000. in Stellar Astrophysics, p. 127, ed. K. S. Cheng (Dordrecht: Kluwer). 
\bibitem[]{}
Lai, D., Chernoff, D.F., \& Cordes, J.M., 2001. {\it Astrophys.~J.}, {\bf 549}, 1111.
\bibitem[Liebend\"{o}rfer et al.~(2001)]{lieben20012} 
Liebend\"{o}rfer, M., Mezzacappa, A., Thielemann, F.-K., 2001. {\it Phys. Rev. D}, {\bf 63}, 104003.
\bibitem[{{Livne}(1993)}]{livne:93}
Livne, E., 1993.  {\it Astrophys.~J.}, {\bf 412}, 634.
\bibitem[]{}
Lyne, A.G., \& Lorimer, D.R., 1994. {\it Nature}, {\bf 369}, 127.
\bibitem[]{}
Ott, C.D., Burrows, A., Livne, E., \& Walder, R., 2003.
``Gravitational Waves from Axisymmetric, Rotating Stellar Core Collapse,"
accepted to the {\it Astrophys.~J.} (astro-ph/0307472).
\bibitem[Rampp \& Janka (2000)]{rampp2000} 
Rampp, M. \& Janka, H.-Th., 2000.  {\it Astrophys.~J.}, {\bf 539}, 33.
\bibitem[]{}
Scheck, L., Plewa, T., Janka, H.-Th., Kifonidis, K., \& M\"uller, E., 2003. astro-ph/0307352.
\bibitem[Thompson, Burrows, \& Pinto (2001)]{thompson3}
Thompson, T.A., Burrows, A., \& Pinto, P.A., 2003.
{\it Astrophys.~J.}, {\bf 592}, 434.
\bibitem[Woosley \& Weaver (1995)]{woosley_weaver} 
Woosley, S.E. \& Weaver, T.A., 1995. {\it Astrophys.~J. Suppl.}, {\bf 101}, 181.
\bibitem[{Zwerger and M{\"u}ller(1997)}]{zm:97}
Zwerger, T. and M{\"u}ller, E., 1997. {\it Astron. and Astrophys.}, {\bf 320}, 209.

\end{thereferences}
\end{document}